\documentclass[twocolumn,aps,floats,superscriptaddress,prl]{revtex4}
\usepackage{epsfig}
\usepackage{amsmath}
\usepackage{graphicx}

\newcommand{\mr}[1]{{{\mathrm{#1}}}}
\newcommand{\mcal}[1]{{\mathcal{#1}}}

\newcommand{\inte}{\int_0^\beta \!\!\!\! \mr{d}\tau}
\newcommand{\w}{\omega}
\newcommand{\s}{\sigma}

\begin{document}

\title{Nanoscale Dynamical Mean-Field Theory for molecules and mesoscopic devices in the strong correlation regime}

\author{Serge Florens}
\affiliation{Institut NEEL, CNRS \& Universit\'e Joseph Fourier, BP 166, 38042
Grenoble Cedex 09, France}

\begin{abstract}
We develop a nanoscale dynamical mean field theory (nano-DMFT) to deal with strong Coulomb 
interaction effects in physical systems that are intermediate in size between 
atoms and bulk materials, taking into account the tunneling into nearby electrodes.
Focusing on a simplified tree-like geometry, the usual DMFT loop simply stops
when the finite lattice is fully covered, starting with an initial seed
provided by the electronic environment at the boundary.
To illustrate this nano-DMFT, we investigate the disappearance of the quasi-particle 
weight in a correlated nano-object near the Mott transition.
In contrast to thermally-driven classical phase transitions, quantum effects
lead to unexpected oscillations of the order parameter, related to the interference of 
coherent renormalized quasiparticles. This behavior also implies a spatially inhomogeneous Mott 
localization process at the nanoscale.

\end{abstract}
\maketitle

The recent developments in the field of strongly correlated electronic systems
are reaching today an interesting point of convergence.
Coming from a macroscopic description of bulk condensed matter, powerful many body 
techniques~\cite{dmft,bosonization}, possibly in combination with ab-initio methods, 
have been tailored to understand an ever growing number of new compounds with surprising 
properties~\cite{imada}.
From a bottom-up perspective, the recent experimental progresses in designing artificial
atoms (quantum dots) or manipulating single atoms and molecules that are
probed by an electronic environment (electrodes) have been met with as well an 
impressive success~\cite{revival}.

How those two fields will merge together is a challenging question raised by today's research, 
whether in semiconducting devices with more complex lithographic patterns, or by the study of 
ultrasmall chunks of correlated materials in quantum transport experiments. 
Such devices ask however for the development of new theoretical methods as well, 
since the inclusion of correlation effects in transport calculations by modern numerical methods can 
only be achieved at the level of single atomic impurities, using Wilson's Numerical Renormalization 
Group method (NRG)~\cite{nrg_wilson,nrg_review}, or simplified linear ``molecules'', by the
Density Matrix Renormalization Group~\cite{dmrg_white,dmrg_nano}, but fails for higher-coordination 
entities. 

In the case of correlated materials in the thermodynamic limit, a philosophy based on
a {\it local effective} description has proved nevertheless very successful. Indeed 
the so-called dynamical mean field theory (DMFT) exactly maps (in the large
coordination limit) a macroscopic lattice problem of correlated fermions onto a single 
atomic level that feels the average effect of the other electrons moving nearby~\cite{dmft,mapping}. 
In this case, the above-mentioned
numerical techniques can be used to solve this simplified effective problem, and allow to capture
quantitatively the effect of the Coulomb interaction on the low energy electronic properties: 
strong signatures indeed appear in the electrical conductivity~\cite{patrice,pbIPT} as electrons start to
localize near the Mott insulating state~\cite{dmft}.

The purpose of this paper is to develop a nano-DMFT approach for correlated nano-objects which
are intermediate in size between the atom and the solid. In the simplest case of
tree-like structures, this can be easily formulated using the cavity
construction~\cite{dmft}, and can be simply described by a 
standard self-consistency DMFT loop which is stopped at a {\it finite} number of iterations, 
related to the linear size of the nano object.
This new insight will be applied both to the Ising and Hubbard models respectively, 
allowing to underline some important physical differences in finite-size many-body 
effects between the classical and the quantum world. As an illustration for the latter, 
we will examine how low energy properties cross over from the single atomic impurity to the 
correlated solid as a function of the system size, focusing on the local quasi-particle weight,
a quantity that characterizes the proportion of electrons that survive the
Coulomb repulsion at low energy, and can serve as an order parameter
for the Mott transition. We find out a surprising oscillatory behavior of this
quasiparticle residue as a function of physical parameters, which reveals the interference 
of coherent quasiparticles in finite size systems. The Mott localization also
happens to be non homogeneous at the mesoscale, with a proportion of sites
considerably ``hotter'' than the others (in the sense of scattering properties).

\begin{figure}[ht]
\epsfig{figure=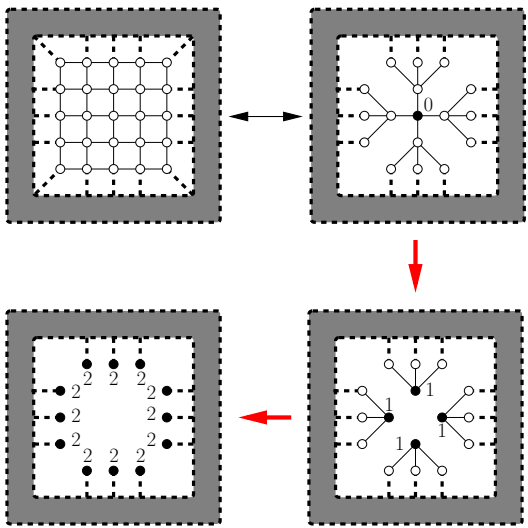,width=7.0cm}
\caption{Correlated nano object of linear extension $2L+1=5$ and coordination number $z=4$, 
defined by a finite piece of either the square lattice (top left) or the Bethe lattice (top right). 
For the latter, which allows the simplest calculations, the nano-DMFT procedure is outlined by the
two further stages of a cavity method given in the lower pictures. The external environment, to which
the nano object couples, is pictured by the shaded area. The nano-DMFT becomes
exact in the limit $z\rightarrow\infty$ keeping $L$ finite.}
\label{cavite}
\end{figure}

We begin by a general description of the cavity method for finite size lattices
in the large coordination limit, see figure~\ref{cavite}.
Although a derivation of the DMFT equations can be done in principle for an arbitrary geometry of the
nano-object, we will consider for simplicity a Bethe lattice of linear size $L$ 
(determined from its center) and coordination $z$; this corresponds to the upper 
right panel of figure~\ref{cavite} for $L=2$ and $z=4$. Focusing on the properties at 
the center of the lattice (site $0$), these can be computed from 
an effective local action involving the site $0$ only, provided one can obtain 
the dynamics at the nearby sites $1$ in the lattice where the site $0$ has been
removed (cavity), see the lower right panel of figure~\ref{cavite}. 
This statement is exact in the limit where the coordination
becomes large~\cite{mapping,metzner}, but can be taken as a local mean field ansatz otherwise.
Similarly, the next shell can also be tackled using a action local to the
sites $1$, given by an effective field provided by the sites $2$ (in the cavity
where the sites of type $0$ and $1$ have been removed).
This is repeated recursively until the boundary is reached, which amounts to an
impurity coupled to a {\it given} external environment (lower left panel of
figure~\ref{cavite}). 

\begin{figure}[ht]
\epsfig{figure=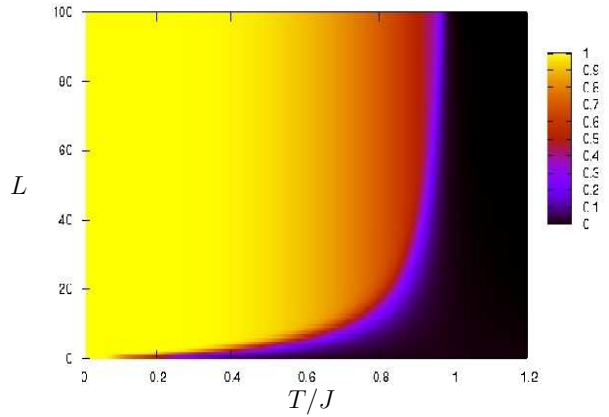,width=8.0cm}
\caption{Average magnetization $\big<S_0\big>$ at the center site for a small
boundary field $h/J=0.01$ as a function of temperature $T/J$ (abscissa) and
number of sites $L$ (ordinate).}
\label{ising}
\end{figure}

Let us put the reasoning backwards, focusing for the sake of clarity on the simple 
Ising model $H = -J\sum_{i,j} S_i S_j - \sum_i h_i S_i$, with a local
external magnetic field that couples to the outer shell, $h_L=h\neq0$ only (due
to the tree structure of the lattice, all those sites are independent, so that
an extra label is not required). These boundary spins are thus free, i.e.
$H^{(L)} = - h S_L$, which gives the 
thermal average $\big<S_L\big> = \tanh(h/T)$. This corresponds to the lower left 
picture in figure~\ref{cavite} for $L=2$. Now the physics of the sites $L-1$ can be obtained 
by an effective impurity problem that reads $H^{(L-1)} = - J\big<S_L\big> S_{L-1}$
(this is exact as $z\rightarrow\infty$), giving $\big<S_{L-1}\big> = 
\tanh(J\big<S_L\big>/T)$. This equation can be used recursively, until 
the center (site $0$) is reached, providing the results seen on figure~\ref{ising}.
Besides the expected monotonous disappearance of the magnetization $\big<S_0\big>$ at
large system size near the critical temperature $T_c/J=1$, we remark that the
thermodynamic limit is only obtained for rather large system size (which
increases with decreasing external magnetic field at the boundary).

We now turn to the more interesting quantum case, considering a Hubbard-Anderson
model~\cite{dmrg_nano} for electrons 
\begin{eqnarray}
\nonumber
H & = & - t \sum_{ij,\s} d^\dagger_{i\s} d^{\phantom{\dagger}}_{j\s}
- \mu \sum_{i\s} d^\dagger_{i\s} d^{\phantom{\dagger}}_{i\s} 
+ U \sum_i d^\dagger_{i\uparrow} d^{\phantom{\dagger}}_{i\uparrow}
d^\dagger_{i\downarrow} d^{\phantom{\dagger}}_{i\downarrow} \\
& & + V \sum_{k\s} [d^\dagger_{L\s} c^{\phantom{\dagger}}_{k\s}
+\mr{h.c.}] 
+ \sum_{k\s} \epsilon_k c^\dagger_{k\s} c^{\phantom{\dagger}}_{k\s}
\end{eqnarray}
with hopping $t$, chemical potential $\mu$, local Coulomb repulsion $U$ within
the correlated lattice, while the hybridization $V$ couples only the outer shell
of sites $L$ (an unnecessary extra index labelling the equivalent sites of type $L$
has been omitted above) to an external electronic bath, which models the electrodes in a
transport experiment.
Following the previous philosophy, we start at the boundary (sites $L$) with single
Anderson impurities (decoupled from the other sites with $i<L$), so that the
local Green function $G^{(L)}(\w)$ is obtained by solving a local effective
action which is nothing but the single impurity Anderson model:
\begin{eqnarray}
\nonumber
\mcal{S}^{(L)} & = & \frac{1}{\beta} \sum_{n\s} [-i\w_n-\mu+\Delta^{(L)}(i\w_n)] 
d^\dagger_{L\s} d^{\phantom{\dagger}}_{L\s} \\
\label{anderson}
& & + \inte \; U d^\dagger_{L\uparrow} d^{\phantom{\dagger}}_{L\uparrow}
d^\dagger_{L\downarrow} d^{\phantom{\dagger}}_{L\downarrow} \\
\label{boundary}
\Delta^{(L)}(\w_n) & = & \sum_k \frac{V^2}{i\w_n-\epsilon_k}
\end{eqnarray}
where $\beta=1/T$, $\w_n = (2n+1)\pi T$. Equation~(\ref{boundary}) provides
thus the boundary condition (and {\it not} an effective Weiss field at this
first stage), similar to the magnetic field $h$ applied at the boundary in the
Ising model. Note that in the absence of an external electronic environment,
$\Delta^{(L)}=0$, the boundary problem reduces to an atomic limit. The
nano-DMFT for the full isolated molecule could then allow the calculation of its 
many-body level spectrum.

Now clearly the effective local action of the sites $L-1$ takes a
form similar to~(\ref{anderson}), with the simple effective bath:
\begin{equation}
\Delta^{(L-1)}(\w_n) = t^2 G^{(L)}(i\w_n)
\end{equation}
due to the Bethe lattice structure (this is obtained by integrating exactly the
sites $L$ in a cavity system constituted of sites $L-1$ and $L$ only, see the
lower right panel of figure~\ref{cavite}). 
By repeating this procedure up to the center of the lattice, we see that 
{\it the local Weiss field Green function $\Delta^{(0)}$ at the center site $0$ in the Bethe 
lattice of linear size $2L+1$ is determined by the $L^{\mr{th}}$ DMFT iteration loop with an 
initial condition~(\ref{boundary}) given by the external electronic environment.}

We now want to investigate some physical results that can be obtained from
this nano-DMFT approach. For simplicity, we focus on the case where the probing
electrodes are modelled by a constant density of states $\rho_0$, so that
the external environment is characterized by a single width parameter
$\Gamma=\pi\rho_0V^2$, providing the starting Weiss field of the DMFT loop
$\Delta^{(L)}(\w_n) = -i\Gamma \mr{sign(\w_n)}$. This is then iterated up to a fixed
number $L$ of times, corresponding to a given linear size $L$ of the lattice, {\it for each value} 
of the Coulomb interaction $U$ and temperature $T$ (the chemical potential is chosen 
as $\mu=U/2$ to stay at particle-hole symmetry).
In contrast, the usual DMFT procedure for a macroscopic system is done until a
small and fixed convergence threshold is reached (including a weighting procedure
that speeds up this convergence), and does not depend on the boundary condition
(unless one considers the first-order metal/insulator transition regime where 
two solutions coexist, see below).

\begin{figure}[ht]
\epsfig{figure=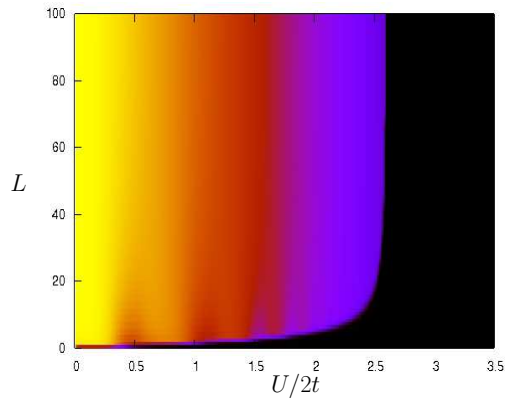,width=7.5cm}
\caption{Quasiparticule weight $Z^{(0)}$ in the center of the molecule as a function of 
Coulomb interaction $U/2t$ (abscissa) and system size $L$ (ordinate), with external 
coupling to the bath $\Gamma/2t=0.001$ and inverse temperature $\beta=800$.}
\label{Z}
\end{figure}

Since the Hubbard model in its thermodynamics limit displays a quantum phase 
transition at increasing $U$ from a correlated metal to a paramagnetic Mott insulator,
we would like to look at a quantity which determines a simple order parameter. 
Focusing on the site in the center of the small island, this is provided by the 
local quasiparticle residue $Z^{(0)}$ at this site $0$, defined as:
\begin{equation}
Z^{(0)} = \frac{1}{1+i\partial \Sigma^{(0)}(i\w)/\partial \w_{|\w=0}}
\end{equation}
introducing the local self-energy $\Sigma^{(0)}(i\w) = i\w +\mu - \Delta^{(0)}(i\w) 
- 1/G^{(0)}(i\w)$.
For the infinite size Hubbard model, the quasiparticle weight is uniform,
decreases from the value $1$ at $U=0$, and vanishes continuously at a critical
strength $U_{c2}/2t \simeq 3$. For $U<U_{c2}$, strongly renormalized
quasiparticules with a reduced coherence time govern the low energy properties.
In parallel to the stabilization of a finite magnetization in the Ising model with a 
non-zero boundary magnetic field, the coupling $\Gamma$ of the outer sites to the 
electronic environment 
plays the role of a dispersion, and acts in stabilizing a
conducting state, even in a finite size system. The quantity $Z^{(0)}$ can
thus be defined for all system sizes and interactions (even above $U_{c2}$ provided
the thermodynamic limit is not reached). We are thus interested in the
convergence of $Z^{(0)}$ with the system size $L$ by increasing the Coulomb
interaction $U$. This is seen on figure~\ref{Z}, which is obtained by using the
iterated perturbation theory (IPT) method~\cite{dmft}.

Superficially, this plot is very similar to the magnetization curve
of the finite size Ising model of figure~\ref{ising}: for any size $L$, the
quasiparticle weight $Z^{(0)}$ drops from $1$ as $U$ increases; also similar to 
the Ising case, the convergence with $L$ towards the thermodynamic result is slow
when the coupling at the boundary to the environment is small. For this precise
plot, this occurs roughly for $L\sim50$, which for an actual Bethe lattice with
connectivity $z=4$, amounts to a macroscopic system with $N=3^{50}$ sites, an
unphysically too large size, due to the fact that the Bethe lattice is very
hollow. In the more realistic case of a square lattice, one obtains a number
of sites $N=100^2$, which is still very challenging for any
direct numerical treatment of the complete many-body problem for such a molecule, providing 
a strong incentive in favor of the more economical nano-DMFT approach. 

There are however some important differences if one compares closely the
classical result, Fig.~\ref{ising}, to the quantum one, Fig.~\ref{Z}.
First, one sees that $Z^{(0)}$ does not actually continuously vanishes
at $L$ large, but presents a small jump at $U_{c1}/(2t)\simeq 2.5$. 
This is due to the existence in the range $U_{c1} < U < U_{c2}$ of
both metallic and insulating solutions, to either of which the system 
converges depending on the initial condition. For the small coupling $\Gamma$
considered here (corresponding to a tiny Kondo temperature
for the initial outer sites $L$), one starts very close to an insulating seed, 
explaining the discontinuous character of this transition seen in figure~\ref{Z}
(the metallic branch is nevertheless correctly reached for $U<U_{c1}$ and
infinite DMFT iterations). One can check that
the true vanishing of $Z^{(0)}$ in the thermodynamic limit at $U_{c2}$ can be 
achieved by increasing the ratio $\Gamma/2t$.

\begin{figure}[ht]
\epsfig{figure=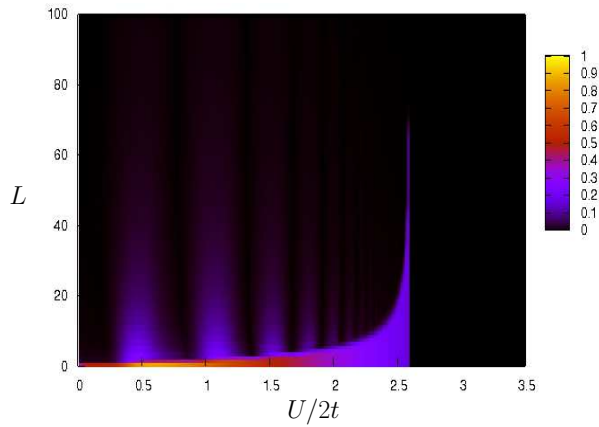,width=8.0cm}
\caption{Relative quasiparticle weight $\delta Z^{(0)} = 
Z^{(0)}_{L\rightarrow\infty}-Z^{(0)}$ for the same parameters as in
figure~\ref{Z}.}
\label{diffZ}
\end{figure}

More interestingly, we discover by closer inspection of figure~\ref{Z}
some small oscillatory behavior of the quasiparticle weigth $Z^{(0)}$
as a function of $U$ for intermediate $L$ sizes (beyond the transient regime of
very small $L$). To display this effect more clearly, we plot on figure~\ref{diffZ} 
the relative weight to the thermodynamic limit $\delta Z^{(0)} = 
Z^{(0)}_{L\rightarrow\infty}-Z^{(0)}$. These oscillations constitute a very surprising 
result that signals that localization (at site $0$) is slightly non-monotonous with
increasing interaction, up to again very large system sizes.
It can be seen also that those oscillations are displaced when comparing two
adjacent sites, thus implying a non-uniform Mott localization process. Note that
a similar behavior appears in the static properties of the Wigner localization in quantum
dots~\cite{ghosal}.
The origin for these effects is the interplay between the interaction induced
localization and the interferences within the nano object (due the presence of
the boundary).
This interpretation is consistent with the observation that these oscillations 
disappear both by increasing temperature,
or by increasing the transmission at the boundary (i.e. making $\Gamma$
comparable to the hopping $t$). This illustrates nicely how correlation effects 
in intermediate size nano objects may lead to unexpected and complex physical behavior. 

We end the paper with some general remarks and desirable extensions of
the nano-DMFT method. We would like first to emphasize that, despite some
resemblances, the nano-DMFT is {\it not} related to the cluster-DMFT 
scheme~\cite{cluster}.
Indeed, in the latter one tries to compensate the sheer locality of DMFT by exactly solving
a finite size plaquette (with a total number $N$ of sites certainly
smaller than 10) embedded in a self-consistent medium that describes the effect
of the remaining sites in an infinite lattice (it is a thus a cavity construction with
a cluster of several sites taken as cavity object). 
The nano-DMFT allows however to reach very large plaquette sizes, typical of big
molecules, at the expense of neglecting some non-local aspects of the physics.
%
%
Further extensions of the present work should focus on investigating transport,
as non homogeneities in the quasiparticle weight will also appear in the scattering 
rate for electron-electron interactions $\tau^{-1}_{e-e} \propto (T/Z)^2$. 
This asks for more sophisticated solvers for the effective 
action~(\ref{anderson}) than the IPT method~\cite{pbIPT}, e.g. with NRG~\cite{nrg_review},
and the development of a consistent formalism for computing the tunneling
conductance. We stress finally that the hypothesis of a Bethe lattice is not central 
for the implementation of the nano-DMFT, and arbitrary ``molecular'' structures can be studied 
along the lines of Ref.~\cite{dmft,dobro,fleck}. This however needs to consider a fully iterated 
self-consistent nano-DMFT loop for the local Green functions, with as many local 
self-energies as the number of sites in the molecule.

To conclude, we have extended the DMFT of macroscopic correlated
fermions models towards a nano-DMFT for generic nanoscale devices 
such as large molecules, quantum dot arrays, or nanostructures implanted
with magnetic impurities. The method is naturally able to capture both
strong electronic interactions and the coupling to an electronic environment.
We have shown that these effects combine in mesoscopic systems to give interesting 
interplays in the Mott localization process that reveals the truly quantum nature 
of this phenomenon.

\acknowledgments{The author thanks V. Dobrosavljevi\'c, A. Lichtenstein and D.
Ullmo for valuable discussions.}


\begin{thebibliography}{5}
\bibitem{dmft} A. Georges, G. Kotliar, W. Krauth and M. Rozenberg, Rev. Mod.
Phys. {\bf 68}, 13 (1996).
\bibitem{bosonization} T. Giamarchi, {\it Quantum physics in one dimension}
(Oxford University Press, Oxford, 2004).
\bibitem{imada} M. Imada, A. Fujimori and Y. Tokura, Rev. Mod. Phys. {\bf 70}, 
1039 (1998).
\bibitem{revival} L. Kouwenhoven and L. Glazman, Physics World {\bf 14}, 33
(2001).
\bibitem{nrg_wilson} K. G. Wilson, Rev. Mod. Phys. {\bf 47}, 773 (1975).
\bibitem{nrg_review} R. Bulla, T. Costi and T. Pruschke, preprint
{\tt cond-mat/0701105}.
\bibitem{dmrg_white} S. R. White, Phys. Rev. Lett. {\bf 69}, 2863 (1992).
\bibitem{dmrg_nano} D. Bohr, P. Schmitteckert and P. Woelfle,
Europhys. Lett. {\bf 73}, 246 (2006).
\bibitem{mapping} A. Georges and G. Kotliar, Phys. Rev. B {\bf 45}, 6479 (1992).
\bibitem{metzner} W. Metzner and D. Vollhardt, Phys. Rev. Lett {\bf 62}, 324
(1989).
\bibitem{patrice} P. Limelette {\it et al.}, Phys. Rev. Lett {\bf 91}, 016401
(2003).
\bibitem{pbIPT} A. Georges, S. Florens and T. A. Costi,
Journal de Physique IV {\bf 114}, 165 (2004).
\bibitem{ghosal} A. Ghosal, A. D. Guclu, C. J. Umrigar, D. Ullmo, and
H. U. Baranger, Nature Physics {\bf 2}, 336 (2006).
\bibitem{cluster} O. Parcollet, G. Biroli and G. Kotliar, Phys. Rev. Lett. 
{\bf 92}, 226402 (2004).
\bibitem{dobro} M. C. O. Aguiar, E. Miranda and V. Dobrosavljevi\'c, Phys. Rev.
B {\bf 68}, 125104 (2003).
\bibitem{fleck} M. Fleck, A. I. Lichtenstein, E. Pavarini, and A. M. Oles,
Phys. Rev. Lett. {\bf 84}, 4962 (2000).ś
 
\end{thebibliography}
\end{document}